\documentclass[manuscript,screen]{acmart}

\usepackage{algorithm}
\usepackage{algpseudocode}
\usepackage{graphicx}
\usepackage{times}
\usepackage{url}
\usepackage[inline]{enumitem}
\usepackage{acronym} 
\setcopyright{none}
\newcounter{todocnt}

\acrodef{RL}{Reinforcement Learning}
\acrodef{DRL}{Deep Reinforcement Learning}
\acrodef{IRL}{Inverse Reinforcement Learning}
\acrodef{SERP}{search engine result page}
\acrodef{IR}{Information Retrieval}
\acrodef{MDP}{Markov Decision Process}
\acrodef{IA}{Interactive Agent}
\acrodef{NLU}{Natural Language Understanding}
\acrodef{ICS}{Intent Completion Score}
\acrodef{IR}{Information Retrieval}
\acrodef{QPP}{Query Performance Prediction}
\acrodef{CIS}{Completion Intent Score}
\acrodef{HCI}{Human Computer Interaction}
\acrodef{SCQ}{Similarity of Collection and Query}
\acrodef{SCS}{Simplified Clarity Score}
\acrodef{CC}{Closeness Centrality}

\begin{document}
%



\title{Making Large Language Models Interactive: A Pioneer Study on Supporting Complex Information-Seeking Tasks with Implicit Constraints}

\author{Ali Ahmadvand}
\email{ali.ahmadvand.66@gmail.com	}
\affiliation{%
 \institution{Emory University	}
  \country{USA}
}

\author{Negar Arabzadeh}
\email{narabzad@uwaterloo.ca}

\affiliation{%
  \institution{University of Waterloo}
  \country{Canada}}

\author{Julia Kiseleva}
\email{julia.kiseleva@microsoft.com	}
\affiliation{%
  \institution{Microsoft Research}
  \country{USA}
}

\author{Patricio Figueroa Sanz}
\email{sefig@microsoft.com	}
\affiliation{%
  \institution{Microsoft}
 \country{USA}}

\author{Xin Deng}
\email{xinde@microsoft.com	}
\affiliation{%
  \institution{Microsoft}
 \country{USA}}

\author{Sujay Jauhar}
\email{sjauhar@microsoft.com	}
\affiliation{%
  \institution{Microsoft Research}
 \country{USA}}

\author{Michael Gamon}
\email{mgamon@microsoft.com	}
\affiliation{%
  \institution{Microsoft Research}
 \country{USA}}
 
\author{Eugene Agichtein}
\email{eugene.agichtein@emory.edu	}
\affiliation{%
 \institution{Emory University	}
  \country{USA}
}

\author{Ned Friend}
\email{Ned.Friend@microsoft.com	}
\affiliation{%
  \institution{Microsoft}
 \country{USA}}

\author{Aniruddha Kulkarni}
\email{anikulk@microsoft.com	}
\affiliation{%
  \institution{Microsoft}
 \country{USA}}

\author{Ahmed Awadallah}
\email{hassanam@microsoft.com}
\affiliation{%
  \institution{Microsoft Research}
 \country{USA}}

\author{Ryen W. White}
\email{ryenw@microsoft.com}
\affiliation{%
  \institution{Microsoft Research}
 \country{USA}}
\begin{abstract}
 
Current interactive systems with natural language interfaces lack the ability to understand a complex information-seeking request which expresses several implicit constraints at once, and there is no prior information about user preferences, e.g.,
\emph{`find hiking trails around San Francisco which are accessible with toddlers and have beautiful scenery in summer’}, where output is a list of possible suggestions for users to start their exploration. In such scenarios, user requests can be issued in one shot in the form of a complex and long query, unlike \textbf{conversational} and \textbf{exploratory} search models, where require short utterances or queries are often presented to the system step by step. This advancement provides the final user more flexibility and precision in expressing their intent through the search process, as well as greater efficiency in their interactions with the system.
Such systems are inherently helpful for day-to-day user tasks requiring planning that are usually time-consuming, sometimes tricky, and cognitively taxing. We have designed and deployed a platform to collect the data from approaching such complex interactive systems.

Moreover, with the current advancement of generative language models such as GPT-based models, understanding complex user requests becomes more possible, however, these models suffer from hallucination in providing accurate factual knowledge. All language models are mostly trained in large part on web-scraped data from the past, which usually is not useful for immediate users' needs.

In this article, we propose an \acf{IA} that leverages Large Language Models (LLM) for complex request understanding and makes it interactive using Reinforcement learning (RL) that allows intricately refine user requests by making them complete, which should lead to better retrieval and reduce LLMs hallucination problems for current user needs. To demonstrate the performance of the proposed modeling paradigm,
we have adopted various pre-retrieval metrics that capture the extent to which guided interactions with our system yield better retrieval results. Through extensive experimentation, we demonstrated that our method significantly outperforms several robust baselines.

\end{abstract}



\maketitle              

\section{Introduction}
\label{sec:intro}
%
%
Recent advances in the field of \ac{NLU}~\citep{devlin2018bert, adiwardana2020towards, brown2020language, raffel2020exploring} have enabled natural language interfaces to help users find information beyond what typical search engines provide, through systems such as open domain and task-oriented dialogue engines~\citep{leiter2023chatgpt,li2018dialogue,li2020guided} and conversational recommenders~\citep{christakopoulou2016towards}, among others.
However, most traditional systems still present with one or both of the following limitations: (1)~answers are typically constrained to relatively simple and primarily factoid-style requests in natural language~\cite{kwiatkowski2019natural,DBLP:conf/eacl/SoleimaniMW21}, as is the case with search engines; and (2)~a requirement on eliciting user preference by asking direct questions about attributes~\citep{kostric2021soliciting}. 
%

\begin{figure}[t!]
\centering
\includegraphics[width=0.7\textwidth]{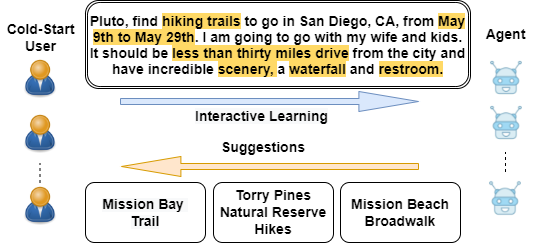}
\caption{An example of a user request expressed complex information needs in natural language, which is processed by \ac{IA} Pluto to retrieve a list of suggestions that at least partially satisfy the specified constraints. The request contains a number of search constraints that are highlighted in yellow.}
\label{fig:running_example}
\end{figure}

However, user information needs, when expressed using natural language, can be inherently complex and contain many interdependent and dependent constraints, as is shown in Figure~\ref{fig:running_example}. When issuing such requests, users may be considered to be in \emph{exploratory mode}; they are looking for suggestions to pick from, rather than a single concrete answer. However, sometimes user preferences can be elicited in multiple known steps if a user is in learning and discovery mode.
The task becomes especially challenging since most real applications~\citep{christakopoulou2018towards} need to support cold-start users~\citep{kiseleva2016beyond, sepliarskaia2018preference}, for whom little to no preferential knowledge is known a priori. This may be due to infrequent visits, rapid changes in user preferences~\citep{bernardi2015continuous, kiseleva2014modelling, kiseleva2015behavioral}, or general privacy-preserving constraints that limit the amount or type of information that can be collected or stored. In this work, we aim to bridge the described gap of processing complex information-seeking requests in natural language from unknown users by developing a new type of application, which will work as illustrated in Figure~\ref{fig:running_example}.
Concretely, our proposed solution is capable of jointly processing complex natural language requests, inferring user preferences, and suggesting new ones for users to explore, given real-time interactions with the \acf{IA}. 

One of the major bottlenecks in tackling the proposed problem of processing complex information-seeking is the lack of an existing interactive system to collect data and observe user interactions. Therefore, we designed a pipeline, which we call Pluto, that allows users to submit complex information-seeking requests. Using Pluto, we leverage human agents in the loop to help users accomplish their informational needs while collecting data on complex search behavior and user interactions~\citep[e.g.,][]{holzinger2016interactive,li2016dialogue}. 

Finally, we propose a novel \ac{IA} that seeks to replace human agents in the loop in order to scale out Pluto to a significantly broader audience while simultaneously making responses a near real-time experience. The proposed \ac{IA} contains a Natural Language Understanding \ac{NLU} unit that extracts a semantic representation of the complex request. It also integrates a novel score that estimates the completeness of a user's intent at each interactive step.
Based on the semantic representation and completion score, the \ac{IA} interacts with users through a \ac{RL} loop that guides them through the process of specifying intents and expressing new ones. The proposed model leverages a user interface to suggest a ranked list of suggested intents that users may not have previously thought about, or even know. Online user feedback is leveraged through these interactions with users to automatically improve and update the reinforcement learner's policies.

Another important aspect we consider is a simple, straightforward evaluation of the proposed approach. We adopt pre-retrieval metrics~\citep[e.g.,][]{sarnikar2014query,roitman2019study} as a means to evaluate the extent to which refinement to the complex request afforded by the \ac{IA} better represents the actual user intent, or narrows down the search space. Our evaluation demonstrates that a better-formulated complex request results in a more reliable and accurate retrieval process. For the retrieval phase, we break down the complex request based on the contained slots and generate a list of queries from the user intent, slots, and location. A search engine API is used to extract relevant documents, after which a GPT-3 \cite{brown2020language} based ranker re-ranks the final results based on the actual slot values or aspects. The final re-ranker considers the user preferences through the aspects values for the slots in the reformulated query.

To summarize, the main contributions of this work are:
\begin{enumerate}[leftmargin=*,label=\textbf{C\arabic*},nosep]
    \item \emph{Designing} a novel interactive platform to collect data for handling complex information-seeking tasks, which enables integration with a human-in-the-loop protocol for initial processing of the user requests and search engines to retrieve relevant suggestions in response to refined user requests (Section~\ref{sec:infrastructure}).
    
    \item \emph{Formalizing new general problem} of interactive intent modeling for retrieving a list of suggestions in response to users' complex information-seeking requests expressed in natural language, such as presented in Figure~\ref{fig:running_example}, where there is no prior information about user preferences (Section~\ref{sec:model}).
    
    \item \emph{Proposing a hybrid model}, which we name \acf{IA}, consisting of an \acf{NLU} and a \acf{RL} component. This model, inspired by conversational agents, encourages and empowers users to explicitly describe their search intents so that they may be more easily satisfied (Section~\ref{sec:solution}).
    
    \item \emph{Suggesting an evaluation metric}, \ac{CIS} that estimates the degree to which intent is expressed completely, at each step. This metric is used to continue the interactive loop so that users can express the maximum preferential information in a minimum number of steps (Section~\ref{sec:eval_metrics}).
\end{enumerate}

\noindent

\section{Background and Related Work}
\label{sec:rel_work}

Our work is relevant to four broad strands of research on multi-armed bandits, search engines, language as an interface for interactive systems, and exploratory search and trails, which we review below.
\subsubsection*{\bf Contextual bandits for recommendation}
Multi-armed bandits are a classical exploration-exploitation framework from \acf{RL}, where the user feedback is available in each iteration~\citep{parapar2021diverse, cortes2018adapting, li2010contextual}. They are becoming popular for online applications such as ranking online advertisements and recommendation systems~\cite[e.g.,][]{ban2021local, joachims2020reveal}, where information about user preferences is unavailable (cold-start users~\citep{bernardi2015continuous,kiseleva2016beyond}) \citep{felicio2017multi}. \citet{parapar2021diverse} proposed a multi-armed bandit model for personalized recommendations by diversifying the user preferences by changing the focus only on past user interactions. Others examined the application of contextual bandit models in healthcare, finance, dynamic pricing, and anomaly detection \citep{bouneffouf2019survey}. Our work adapts contextual bandits paradigm to the new problem of interactive intent modeling for complex information-seeking tasks.
\subsubsection*{\bf Search engines}
Commonly used search engines such as Google and Bing provide platforms focusing on the document retrieval process through search sessions~\cite{Hassan_wsdm_2010, kiseleva2014modelling, kiseleva2015behavioral,ageev_sigir_2011}. Developing retrieval models that can extract the most relevant documents from an extensive collection has been well-studied~\citep{croft2010search} for decades. The developed retrieval models focus on retrieving the most relevant documents corresponding to user intent, represented with textual and contextual information within and across search sessions~\citep{kotov2011modeling}. Although extracting relevant documents is necessary, it is not always sufficient, especially when the users have a complex information-seeking task~\citep{ingwersen2006turn}. 
\subsubsection*{\bf Language as an interface for interactions}
\ac{NLU} have been the important direction for human-computer interaction and information search for decades~\citep{woods1972lunar, codd1974seven, hendrix1978developing}. The recent impressive advances in capabilities of \ac{NLU}~\citep{devlin2018bert, LiuRoberta_2019, clark2020electra, adiwardana2020towards, roller2020recipes, brown2020language} powered by large-scale deep learning and increasing demand for new applications has led to a major resurgence of natural language interfaces in the form of virtual assistants, dialog systems, semantic parsing, and question answering systems~\citep{liu2017iterative, liu2018adversarial, dinan2020second, zhang2019dialogpt}. The scope of natural language interfaces has been significantly expanding from databases~\citep{copestake1990natural} to knowledge  bases~\citep{berant2013semantic}, robots~\citep{tellex2011understanding}, virtual assistants~\citep{kiseleva2016understanding, kiseleva2016predicting}, and other various forms of interaction~\citep{fast2018iris, desai2016program, young2013pomdp}.
Recently, the community has focused on continuous learning through interactions, including systems that 
learn a new task from instructions~\citep{li-etal-2020-interactive}, 
assess their uncertainty~\citep{yao-etal-2019-model}
and ask feedback from humans in case of uncertainty~\citep{aliannejadi2021building, Aliannejadi_convAI3} or for correcting possible mistakes~\citep{elgohary-etal-2020-speak}.

\subsubsection*{\bf Exploratory search, tours, and trails}
Exploratory search refers to an information-seeking process in which the system assists the searcher in understanding the information space for iterative exploration and retrieval of information~\citep{ruotsalo2018interactive, hassan2014supporting, white2008evaluating}. Anomalous states of knowledge (ASKs) \cite{belkin1980anomalous} motivate the need to search and drive demand for search systems. According to the ASK hypothesis, users usually can struggle to conceptualize and formulate their information needs as search queries, which may miss some essential information~\citep{liu2015personalizing, white2009exploratory}. In such cases, the system should assist the user in specifying their intent~\citep{marchionini2006exploratory}. Through a search log analysis, \citet{odijk2015struggling} shows that there are many searches where users may struggle to formulate their search query or they may simply be exploring to learn about a new area. New search interface designs may be required to support searchers through their information-seeking process~\citep{villa2009aspectual}. \emph{Tours and Trails} are another group of tools that were developed to guide users to accomplish search tasks. Guided tours are common in hypertext systems \cite{trigg1988guided} and similar ideas could be applied in the context of search \cite{hassan2012task}. Surfacing common trail destinations in search interfaces can help people find information targets more quickly \cite{white2007studying}. Search engines may also present full trails as a way to explore, learn, and complete multi-step tasks \cite{singla2010studying}. \citet{olston2003scenttrails} proposed ScentTrails that leverage an interface that combines browsing and searching and highlights potentially relevant hyperlinks. WebWatcher~\cite{joachims1997webwatcher}, like ScentTrails, underlined the relevant hyperlinks and improved the model based on the implicit feedback collected during previous tours.

\smallskip

\noindent
To summarize, the \textbf{key distinctions} of our work compared to previous efforts are as follows.
Similar to the exploratory search, trails, and conversational search, our model proposes an iterative information-seeking process and designs an interface for user interactions to guide struggling users and help them better understand the information space. However, that work that only focuses on user interaction modeling and limits users in issuing short and imprecise queries and utterances, our model provides a platform for users to express their information needs in the form of long and complex requests.
Users can utilize this capability to express their intent more accurately and prune significant parts of the search space for the exploratory search process. 
Adding this capability needs an advanced \ac{NLU} step and different machine learning components to understand and guide the final user through the search process. To this end, the proposed system has two new components, an \textbf{intent ontology} and a \textbf{profile} for partitioning the information space, enabling the \ac{IA} to help users be more effective in exploring the search space.

\begin{figure*}[!t]
\centering
\includegraphics[width=300pt]{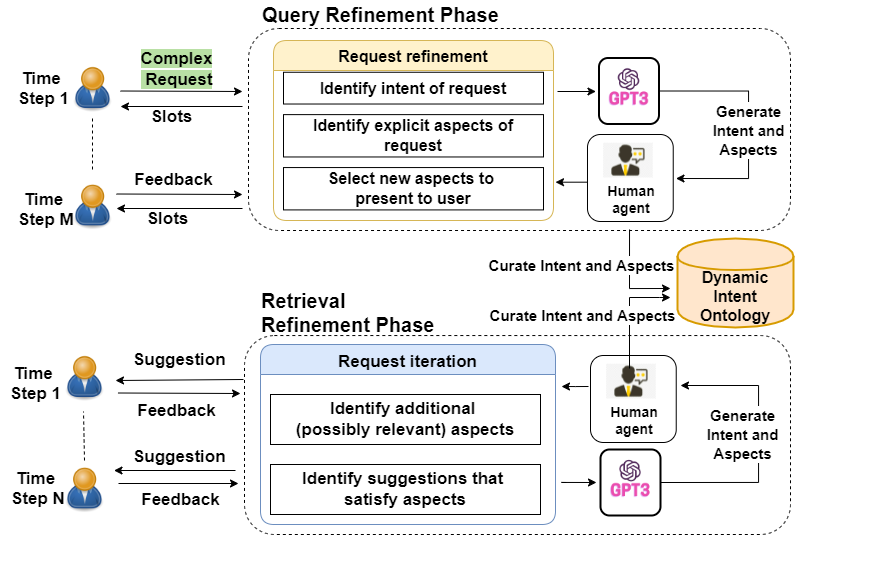}
\caption{\small Description of the Pluto infrastructure, which is used for interactive data collection. Phase 1 describes the refinement of complex requests, and phase 2 depicts the main steps of suggestions retrieval.}
\label{fig:assistantinfrastructure}
\end{figure*}

\section{Pluto: data collection infrastructure }
\label{sec:infrastructure}

Since the proposed problem is novel and requires non-trivial user interaction data, we designed a new pipeline, Pluto, to collect such data. Pluto uses a human-in-the-loop setup for data collection and curation. It is comprised of two main components depicted in Figure~\ref{fig:assistantinfrastructure}:
\begin{enumerate}[leftmargin=*, label=\textbf{Phase\arabic*}, nosep]
    \item \emph{Refinement of complex user request in natural language};
    \item \emph{Refinement of retrieved list of suggestions.}
\end{enumerate}

\subsubsection*{\bf{Complex user request refinement.}}
When a user issues a request in natural language to express their complex information needs, which potentially has many expressed constraints (see Figure~\ref{fig:running_example} for several mentioned in the example), GPT-3~\cite{brown2020language} is leveraged to understand the request's intent and identify explicitly mentioned aspects. 

Once GPT-3 has identified these aspects, they will be used as the initial set for the request. To further expand this set, this phase will proceed to identify an additional list of aspects to be presented to users as a supplemental set of relevant considerations for their request.

As stated, Pluto has integrated human-in-the-loop into its pipeline. The goal of human agents is to intervene at certain stages of the system to offer human judgment. One such intervention occurs when agents review users' requests, at which point they can correct the aspects this phase identified in the request as well as add new ones to better serve user needs.

\subsubsection*{\bf{Suggestion refinement.}}
Here, Pluto performs two tasks. First, it receives the slots selected by the user for processing and suggests additional slots (so as to further narrow down the request, with the aid of the user). These new slots can be generated via GPT-3 or by intervention from the human agents. Second, Pluto leverages the search engine to produce a series of suggestions that meet the slots for the request as well as the new slot proposal. GPT-3 is leveraged at this stage to aid in determining which potential suggestions meet which aspects from the request so that the system can rank them. Human agents then make final decisions on which suggestions to present to users.
Once that is done, users can either accept the suggestions if they are satisfying, or request another iteration of the retrieval phase. When users request another iteration, they may change either the wording of the request or add/remove aspects from it (including the newly suggested ones). Additionally, for any iteration of this phase, users can provide feedback that is captured via a form to help refine the system.

Finally, human agents are responsible for another very valuable and essential contribution: intent and aspect curation. In either of the phases described above, GPT-3 may suggest various aspects and intents that sometimes are not as relevant or useful. All of these are considered entries into the dynamic intent ontology. However, human agents then curate them. Intent and aspects that are considered higher quality by the agents are then given more weight when suggesting aspects in either of the two phases.

\subsubsection*{\bf{Data handling.}} Users of Pluto were supplied with a consent form explaining that their requests and interactions would be viewed by human agents and some members of the development team. Further, the human agents in the loop also consented to have their interactions with the system recorded. All data and interactions were anonymized, and no personal identifiers of users or agents were retained in any of the modeling and experimentation in our work.%

Next, we will describe the formal problem description and elaborate on the problem formulation and the proposed interactive agent.

\section{Problem description}
\label{sec:model}

\label{sec:model_overview}

In this section, we formalize the problem of interactive intent modeling for supporting complex information-seeking tasks.

\begin{figure*}
\centering
\includegraphics[width=0.95\textwidth]{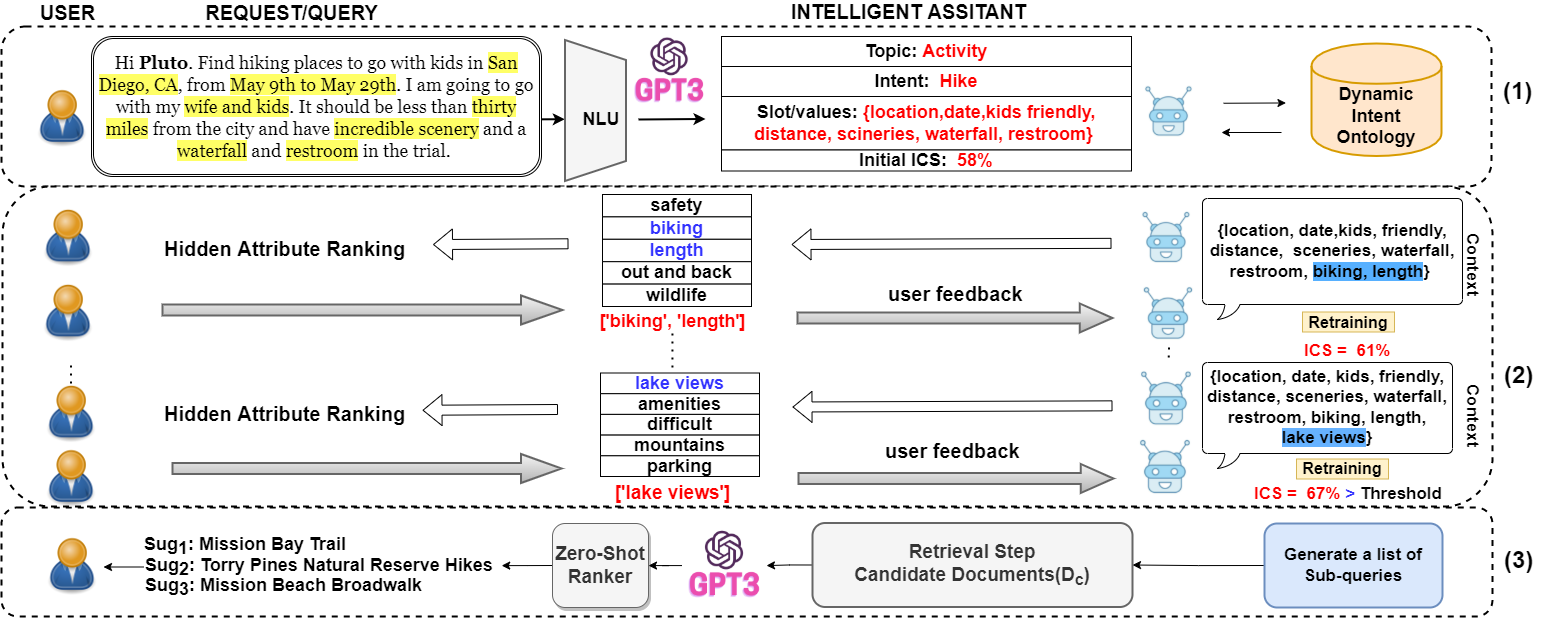}
\caption{\small The proposed \acf{IA} model, where (1)~represents the \ac{natural language understanding} section; (2)~shows the interactive intent modeling via the \ac{reinforcement learning} model; and (3)~is the retrieval component.}
\label{fig:model}
\end{figure*}

\subsubsection*{\bf Notation.}
\label{sec:problem_formulation}

We begin by formally defining used notation as follows: 



\begin{itemize}[leftmargin=*,nosep]

    \item \emph{User request $(cr)$:} is a complex information-seeking task expressed in natural language, which contains multiple functional desiderata, preferences, and conditions (e.g.~Figure~\ref{fig:running_example}).
    
    \item \emph{Request topic $(\tau)$:} determines the topic the request belongs to, e.g. ``activity'' or ``service'', where $\tau \in T$ is the list of all existing topics.

    \item \emph{User intent $(i)$:} for each topic $\tau$, a list of user intents can be defined. Intents are the identification of what a user wants to find. For example, in Figure~\ref{fig:running_example}, the user request has an ``activity'' topic, with a ``hiking'' intent. This definition allows having identical intents in different domains, where $i \in I$ is the list of all intents.

    \item \emph{Slot $(\xi)$ and aspect $(\alpha)$:} for each specific topic $\tau$ and user intent $i$, a list of slots $\xi$ are defined that describe features and properties of the intent $i$ in topic $\tau$, and aspect (values) $\alpha$ is a restriction on the slot $\xi$. For example, from Figure~\ref{fig:running_example}, ``date'' is a slot related to ``hiking'', with aspect value of ``May 9th to May 29th, 2021''.
    
    \item \emph{Intent completion score $(ICS_i)$:} is a score to estimate the completeness of user intent $i$ in the interaction step $j$.
    
    \item \emph{Semantic representation $(\sigma)$:} is an information frame that represents an abstract representation of the $cr$ as $(\tau,i, [ \xi_{0 \dots n} ],ICS_{(\tau,i)} )$   

    \item \emph{Intent ontology $(\Omega)$:} is the graph structure representing relations among the defined domains, intents, and slots.

    \item \emph{Intent profile $(\Phi)$:} is the list of all conditional distributions $P(\xi|i,\tau)$, for slot $\xi$ with respect to topic $\tau$ and intent $i$. It can be changed over time via user interactions with specific intent $i$ and topic $\tau$.
    
    \item \emph{List of retrieved suggestions $S=(Sug_{0},\dots,Sug_n)$:} is the list of retrieved suggestions in response to $cr$.
\end{itemize}



\subsubsection*{\bf Problem formulation.}

\begin{algorithm}
\label{alg:1}
\caption{The proposed interactive user intent modeling for supporting complex information-seeking tasks.}
\label{alg:cap}
\begin{algorithmic}
\State \textbf{Input: } User complex search request
\State
\State \textit{NLU component starts} \Comment{/*This is a comment*/}
\State  Pre-processing of complex request $cr$ \\
\state  \mathrm{F}orming prompt by concatenating $cr$ with training data
\State $\sigma = ( \tau$, $i$, $\xi_{0 \dots n}$, $ICS_{(\tau,i)}  )  \gets$ \textbf{NLU}($cr$)
\State Initialize $c$ = one of the methods in section \ref{sec:interactive}
\State
\State \textit{RL model starts} \Comment{/*This is a comment*/}
\While{$ICS_{(\tau,i)} \leq \mu(P(\xi_k | i, \tau )) + sd(P(\xi_k | i, \tau ))$}
\If{User Feedback}
    \State 1- Retrain the RL policy based on user feedback
    \State 2- Update context $c$ based on user feedback
    \State 3- Update $ICS_i$ in Eq.~\ref{eq:ics}
     \State $ICS_{(\tau,i)} = \Sigma^{n_{\tau,i}}_{k=1}{P(\xi_k | i, \tau )} + \Sigma^{m_{\tau,i}}_{j=1}{P(\xi_j | i, \tau )}$
\Else \hspace{5pt} break
\EndIf
\EndWhile
\State
\State \textit{Retrieval step starts} \Comment{/*This is a comment*/}
\State $p_s$ = [] \Comment{/*potential suggestions*/}
\For{slot $\xi$ in context $c$} 
\State query $\gets i + with + \xi + in + $ \textit{location} 
\State top-10 documents $\gets$ search engine\_API (query) 
\State Update $  p_s \gets  p_s \cup$ top-10 documents 
\EndFor
\State \textbf{return} top-K results from \textbf{GPT-3 Ranker}($P_s$)   
\end{algorithmic}
\end{algorithm}

This section provides a high-level problem formulation. The desired \ac{IA} aims to map a request expressing complex information-seeking task $cr$ to a set of relevant suggestions $S$, as illustrated in Figure~\ref{fig:model}. The proposed model is comprised of three main components:

\begin{enumerate}[leftmargin=*, nosep]
   \item \textbf{\acf{NLU} component:} consists of a topic and intent classifier, and a slot tagger to extract topic $\tau$, user intent $i$, and a list of slots $\xi_{0 \dots n}$, respectively. The unit leverages GPT-3 to improve and generalize the predictions for unseen slots. Finally, NLU generates the semantic representation $\sigma = (\tau,i,[\xi_{0 \dots n}], ICS_{(\tau,i)})$ for a complex request $cr$.
   
   \item \textbf{Interactive intent modeling component:} is an iterative model leveraging contextual multi-armed bandits~\citep{cortes2018adapting} that receives the semantic representation $\sigma$ and context $c$ for the request $cr$ from the NLU unit and predicts the most relevant set of slots for $c$.
   
\item \textbf{Retrieval component:} generates a sequence of sub-queries based on the list of slots and their corresponding aspects. Relevant documents are retrieved from the Web using Search engine API and ranked by GPT-3 to provide the final list of retrieved suggestions $S=\{Sug_0, \dots, Sug_k\}$.
\end{enumerate}

\smallskip
\noindent
To summarize, this section formally defines a problem we intend to solve (Algorithm~1) in the next section. 

\section{Method Description}
\label{sec:solution}

This section presents a detailed description of the proposed strategy to model \acf{IA}. 

\subsection{Creating intent profile $\Phi$}
\label{sec:dynamicontology}

\begin{figure*}
\centering

\includegraphics[width = 350pt]{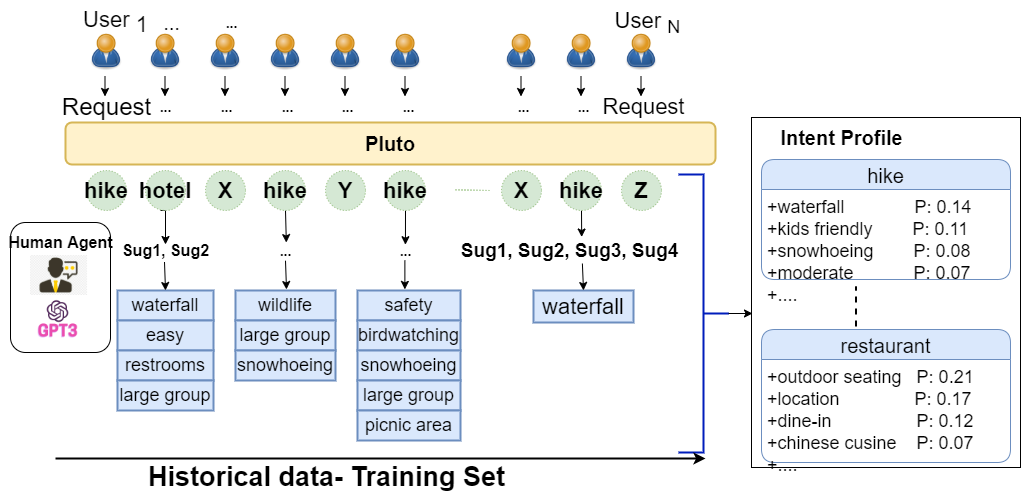}
\caption{\small Intent profile creation through historical user interactions, where $Sug_i$ represents the $i-th$ suggestion with associate slots.}

\label{fig:intentontology}
\end{figure*}

Based on the intent ontology $\Omega$ created in Section~\ref{sec:infrastructure}, and historical users' interactions with topic $\tau$, intent $i$, slot $\xi$, a dynamic intent profile $\Phi$ can be formed as shown in Figure \ref{fig:intentontology}. To do so, for each individual $\xi$, $i$, and $\tau$, the intent profile stores a conditional probability, which can be updated in real-time using new user interactions with triple $(\tau,i,\xi)$. The conditional probability $P(\xi | i,\tau )$ is computed as follows:

\begin{equation}
    P(\xi_k | i, \tau ) = \frac{Frequency(\xi_{(\tau,i,k)})}{\Sigma^{N_\xi}_{j=1} {Frequency(\xi_{(\tau,i,j)})}}
    \label{eq:ontology}
\end{equation}
where $\xi_{(\tau,i,k)}$ is the $kth$ slot for intent $i$, and topic $\tau$, $N_i$ is the number of slots for intent $i$ and topic $\tau$ in intent ontology $\Omega$.

\subsection{NLU component}
\label{sec:nlu}

\begin{figure}
\centering
\includegraphics[width = 300pt]{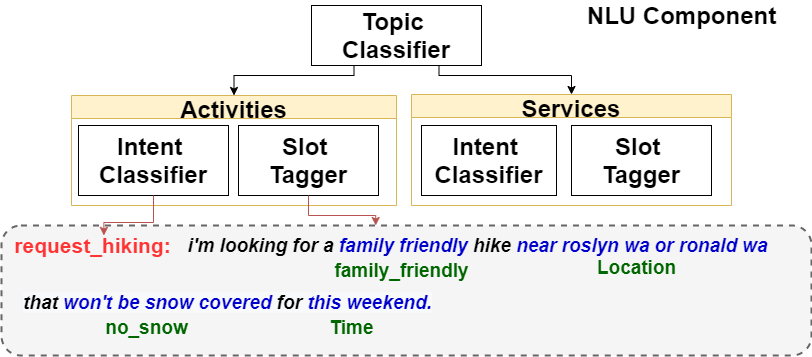}
\caption{\small Flowchart of the proposed IA model.}
\label{fig:nlu}
\end{figure}

The NLU unit contains three main components: (1) a topic classifier, (2) an intent classifier, and (3) a slot tagger. For each incoming complex request $cr$, this unit generates a semantic representation as follows: $\sigma =  (\tau, i, [\xi_{0 \dots n)}],ICS_{(\tau,i)})$. Figure~\ref{fig:nlu} shows the NLU unit for the proposed model.

\subsubsection*{ \bf GPT-3}
To generate the semantic representation, we leveraged GPT-3~\citep{brown2020language}, a generative large language model trained on massive amounts of textual data that has proven capable of natural language generalization and task-agnostic reasoning. One of the hallmarks of GPT-3 is its ability to generate realistic natural language outputs from few or even no training examples (few-shot and zero-shot learning).

The creativity of the model for generating arbitrary linguistic outputs can be controlled using a temperature hyperparameter. We use an internal deployment of GPT-3\footnote{Based on \url{https://beta.openai.com/}.} as the basis for our NLU.

We leveraged the few-shot prompting technique \cite{brown2020language, sun2022recitation} for inference, where the training data collected in section \label{sec:nludatacollection} is used to form the few-shot prompt for all the GPT-3 requests. Finally, the actual request is concatenated with the data and forms the final prompt. 

\subsubsection*{ \bf Intent Completion score}
We propose a score \ac{ICS} to manage the number of interactions for the interactive steps. The ICS value can be calculated using the semantic representation $\sigma$ and the generated dynamic intent profile $\Phi$. The initial ICS value is equal to the summation over all the conditional probabilities of slots in the request. Then, in the following steps, ICS becomes updated by new slots that the user selects.

\begin{equation}
\label{eq:ics}
    ICS_{(\tau,i)} = \Sigma^{n_{\tau,i}}_{k=1}{P(\xi_k | i, \tau )} + \Sigma^{m_{\tau,i}}_{j=1}{P(\xi_j | i, \tau )}
\end{equation}
Where $n_{\tau,i}$ is the number of explicitly \textbf{mentioned} slots in the $cr$ and $m_{\tau,i}$ is the number of \textbf{selected} slots through the interactive steps. Also, $P(\xi | i, \tau )$ indicates the conditional probability extracted from intent profile $\Phi$ in Eq.~\ref{eq:ontology}. 

\subsection{Interactive user intent modeling}
\label{sec:interactive}

\begin{algorithm}
\label{alg:cap}
\caption{Contextual Multi-armed Bandit Model. $CBM_{(\tau,i)}$ is the contextual bandit model trained on the topic $\tau$ and intent $i$, $\Pi_\theta (.|c)$ is the policy to train the $CBM_{(\tau,i)}$ with respect to context $c$.}

\begin{algorithmic}
\State \textbf{Input: } semantic representation: $\sigma = (\tau,i,[\xi_{0 \dots n}],ICS_{(\tau,i)})$
\State Generate context vector $c$ using Method 1,2, or 3 in section \ref{method1}
\State Select the $CBM_{(\tau,i)}$ based on $(\tau, i)$ tuple
\State Initialize Policy $\Pi_\theta (.|c)$ with random weights (a feed-forward Neural Network or a linear regression)
\For{each step and context of $c$ } 
\State Sample actions $a_ {0 \dots k} $ from list of actions: $a_t \sim \Pi_\theta (.|c)$
\State Receive reward $\epsilon$ (user feedback by selecting actions) 
\State Add sampled actions $a_{{0 \dots k}}$ to the list of observed arms
\State Update Policy \textbf{$\Pi_\theta$} with new reward
\State Update $ c \gets c \cup a_s$  \Comment{/*$a_s$ is the selected actions  $\subset a_{{0 \dots k}}$*/}
\EndFor

\end{algorithmic}
\label{alg:cap}

\end{algorithm}

\begin{figure}
\centering

\includegraphics[width = 280pt]{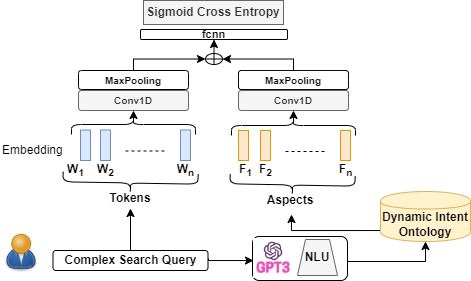}

\caption{ \small Deep neural network for slot suggestion.}

\label{fig:session}
\end{figure}

We leveraged contextual multi-armed bandits to model online user interactions. In each iteration, the system interacts with users, receives user feedback, and updates its policies. Multi-armed bandits \citep{barraza2020introduction} are a type of RL model that make rewards immediately available after the interaction of an agent with the environment. Contextual multi-armed bandits are an extension of multi-armed bandits, where the context of the environment is also modeled in predicting the next step. Contextual multi-armed bandits are utilized in the interactive agent as users are capable of providing feedback for the agent in each step. 
We trained a separate contextual multi-armed bandit to represent each $(\tau, i)$ pair as shown in Algorithm \ref{alg:cap}. The corresponding bandit model is then invoked at the inference time, based on the semantic representation $\sigma$. One of the main elements in designing the contextual bandits is how to represent the context $c$. To this end, we suggested three different methods that are described in the following sections.


\subsubsection*{\bf Method 1:}
\label{method1}

This method uses a one-hot representation of the semantic representation $\sigma$. During the interactions with our agent, the one-hot representation is updated by adding newly selected slots. As a result, the size of the context $c$ equals the number of slots for each specific intent. 

\begin{equation}
    c = \vec{O_k} = \sum^{N}_{j=1} {\bigg\{^{1\hspace{0.5cm} x_{(j,k)} \in \xi_i}_{0\hspace{0.5cm} x_{(j,k)} \notin \xi_i}}
\end{equation}
Where $\vec{O_k}$ is the one-hot vector of the collected slots in the interaction step $k$. $N$ is the total number of slots in the interaction step $k$, and $\xi_i$ is the slots belonging to intent $i$.

\subsubsection*{\bf Method 2}
\label{method2}

In method 2, the request representation is concatenated with the one-hot representation of the slots to enrich the context representation. We used the Google Universal Sentence Encoder(USE) \cite{cer2018universal}, which is trained with a deep averaging network (DAN) encoder for encoding text into a 512-dimensional vector for each request. 

\begin{equation}
\vspace{-0.5em}
    c = \vec{O_k} + \vec{USE_Q}
\end{equation}
Where $\vec{O_k}$ is the one-hot vector of the collected slots at step $k$.

\subsubsection*{\bf Method 3}
\label{sec:popularity}

Inspired by session-based recommender systems~\citep{wu2017session}, we developed a deep learning model in Figure~\ref{fig:session} to extract the slot representations. users were excluded from the model as we only focused on intent modeling independent of the user. The goal is to predict the list of slots most likely to be selected by the user, given the input request and explicitly mentioned lists of slots in semantic representation $\sigma$.

The model consists of (1) an embedding layer, (2) a representation layer, and (3) a prediction layer. We used sigmoid cross-entropy to compute the loss since the task is a multi-label problem: a subset of slots is predicted for an input list of slots and the request representation. Finally, max-pooling is done across all the slot embeddings and concatenated with the request embedding vector to represent $c$.

\begin{equation}
    c =  Max Pooling (\vec{O_k} * \xi_{(\tau,i,j,k)}) +  \vec{USE_Q}
\end{equation}
where $\xi_{(\tau,i,j,k)}$ is the $j^{th}$ slot embedding, with respect to intent $i$ and topic $\tau$, and $\vec{O_k}$ is the one-hot vector of the collected slots in step $k$.

\subsubsection*{\bf Threshold to stop iterations:} We leverage the $ICS_{(\tau,i)}$ score to stop the contextual bandit iterations, which has a steady increase in its value through the interactions. To this end, when this value becomes greater than a threshold, the contextual bandit model stops iteration. The threshold varies per $(\tau, i)$ pair. Hence, we consider a threshold value of $ICS_{(\tau,i)} \leq \mu(P(\xi_k | i, \tau )) + sd(P(\xi_k | i, \tau ))$ the mean plus the standard deviation of the slot distribution within $(\tau, i)$.

\subsection{Retrieval Component}
\label{sec:retrieval}
To extract the final recommendations for the users, we use a retrieval engine that consists of two main components: 1) search retrieval and 2) ranking. For the retrieval part, we need to collect a corpus that is representative of the search space on the Web. Then, we can evaluate the pre-retrieval metrics is discussed in section \ref{sec:eval_metrics}. for both initial requests and reformulated requests at inference time. 

\subsubsection*{\bf Corpus collection:} To generate the corpus, we need to issue a series of queries to a search engine that will capture the search space of the web. Algorithm \ref{alg:subquery}, in section \ref{sec:appendix}, shows the steps we used to generate these queries and collect the corpus. In essence, we leveraged a pool of sub-queries derived from the internal intent ontology. To create these sub-queries, we use the idea of request refinement using request sub-topics \cite{nallapati2006evaluating}, and generated a list of them by combining each selected topic/intent with the set of aspects we have associated with it.

Finally, these queries were issued to the Bing Web Search API, and the top 100 results (consisting of the page's title, URL, and snippet) for each query were added to the corpus.

\subsubsection*{\bf Few-shot Ranker:}
A few-shot GPT-3 model, which has been fine-tuned on a limited number of training samples, is deployed on the pool of potential suggestions extracted from the Web Search API. The GPT-3 ranker then ranks all the potential suggestions concerning the evolved user intent and the actual aspect values $\alpha$. The GPT-3 ranker considers the user preferences for the final ranking results. 

\section{datasets}
\label{sec:datasets}

To evaluate the proposed interactive model, we leveraged the real data collected through user interaction with Pluto. We collected more than $16,699$ user requests with $166,990$ user interactions for training, and $1,140$ user requests with 13,840 interactions for testing.
In Section~\ref{sec:nludatacollection}, we describe a crowd-sourcing procedure that is designed to collect annotated data, which is used to train and test the slot tagger in the \acf{NLU} unit.
Section~\ref{sec:datasetcollection} describes the interactive data collected via Pluto (Section~\ref{sec:infrastructure}). \footnote{The datasets contain potentially sensitive data and cannot be shared publicly due to privacy concerns. However, we believe that using the presented descriptions the dataset collection can be reproduced.}More details about data collection steps and evaluating the annotation process are described in section \ref{sec:nludatacollection} and \ref{sec:datasetcollection}. 

\subsection{Dataset Collected for \ac{NLU} unit}
\label{sec:nludatacollection}

\begin{figure*}
\centering
\includegraphics[width = 280pt]{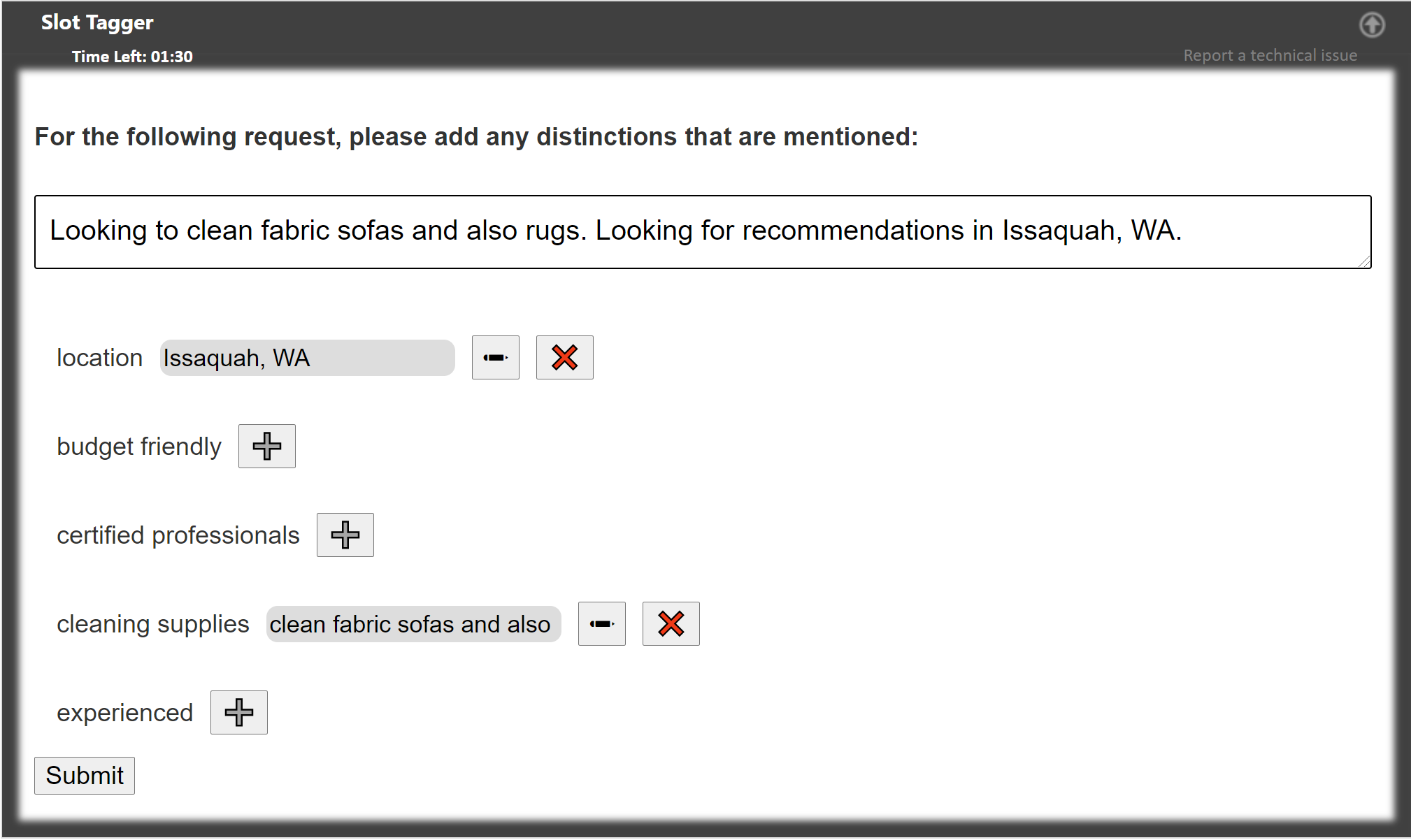}
\caption{ \small Labeling app designed to collect data.}
\label{fig:uhrs}
\end{figure*}

To collect the data for training and evaluating the \ac{NLU} model, we use a crowd-sourcing platform that provides an easy way for researchers to build interfaces for data collection and labeling. Using the platform, we developed a simple interface that presented annotators with a natural language request paired with up to five possible slots. Annotators were then asked to mark relevant slots and given the opportunity to highlight sections of the request that mapped to the slot $\xi$ to their corresponding aspect $\alpha$ in question. Figure~\ref{fig:uhrs} shows a screenshot of the labeling interface.

The set of requests and slots presented to annotators was created from a seed set of $3,246$ requests, where each request was paired with all the slots from the subsuming intent. Three annotators then used the interface to map slots to requests as appropriate. 


\paragraph{Evaluating quality of the collected dataset.}
 Requests were randomly selected from two different topics  and 14 user intents, respectively (Table~\ref{tab:kripendorf}). We only chose two topics as the selected intents were a part of these two topics. Three different human annotators manually labeled these queries through the data collection interface described in the previous section. Table~\ref{tab:kripendorf} presents Krippendorff's alpha scores \cite{krippendorff2011computing} across all the intents. A score above or equal to 0.667 is often considered a good reliability test. The results demonstrate an acceptable agreement among all annotators, except ``Hike'' intent which shows a moderate agreement~\cite{krippendorff2011computing}. After evaluating the $\Omega$, we notice that the slots for ``hike'' intent have overlapped, meaning there are slots that refer to the same thing with different textual representations. These semantic overlaps happened even after normalization with the clustering, which sometimes confuses annotators.

\begin{table}[!t]
\small
\begin{center}
\begin{tabular}{p{0.8cm}|p{1.3cm}|c|c|p{1.3cm}|c}
\bottomrule
\bf Topic &\bf Intent&\bf K (dist)& \bf Topic&\bf Intent&\bf K (dist)\\ 
\bottomrule
Service  & restaurants  & 0.74 \footnotesize(12\%)  & Service& appliance  & 0.71 \footnotesize(11\%)\\   
Service& electrician  &  0.79 \footnotesize(13\%)  & Service& hotel &0.71 \footnotesize(2\%)\\
Service&landscaping  & 0.67  \footnotesize(16\%) & Service& handyman& 0.75 \footnotesize(2\%)\\
Activity& hike  & 0.58   \footnotesize(10\%)& Service& cleaners& 0.69 \footnotesize(4\%)\\
Activity& general  & 0.74 \footnotesize(8\%)  & Service& remodeling& 0.82 \footnotesize(3\%)\\
Activity& spring break  & 1.00 \footnotesize(5\%) & Activity& daytrip& 0.73 \footnotesize(2\%)\\
Activity& campground  & 0.74 \footnotesize(6\%)  & Activity& summercamp& 0.75 \footnotesize(6\%)\\
        \bottomrule
\end{tabular}
\end{center}
\caption{Krippendorff's score across all $(\tau,i)$ tuples in the $\Omega$. Where $K$ and dist stand for Krippendorff's score and the distribution of the tuple in the dateaset.}
\label{tab:kripendorf}
\end{table}

\begin{figure*}
\centering
\includegraphics[width = 280pt]{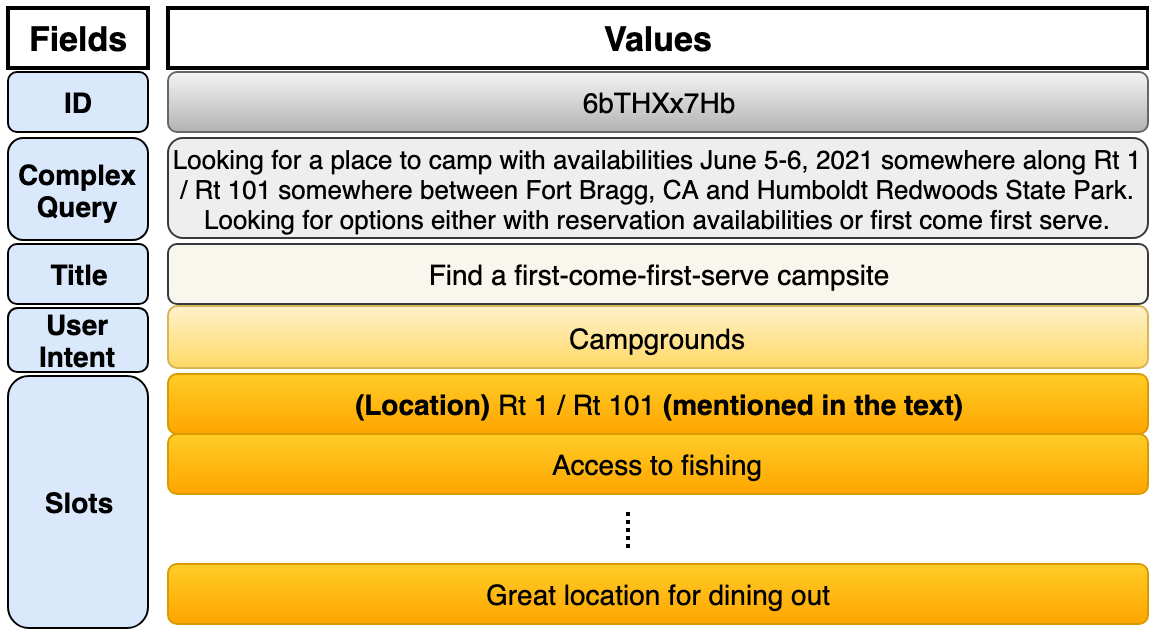}
\caption{\small A data point from Pluto interactive log.}
\label{fig:datapoint}
\end{figure*}

\subsection{Dataset Collected via Pluto}
\label{sec:datasetcollection}
The data for training and evaluating our proposed model is collected during six months of Pluto proprietary interactive logs described in Section~\ref{sec:infrastructure}. We used the first five months to form the training set and reserved the last month for testing. Since GPT-3 is a generative model, the suggested slots during data collection may not be expressed identically, despite representing the same underlying intent (e.g.``access to parking'' and ``parking availability''). To address this issue, we used a universal sentence encoder~\cite{cer2018universal} to softly match a generated slot to a slot in $\Omega$. The slot with the lowest cosine distance is considered the target slot. Figure~\ref{fig:datapoint} illustrates an example of a data instance.

Pluto is capable of covering hundreds of different user intents. In this study, however, we selected the $14$ most frequent search intents in the logs because we observed a sharp drop-off in frequency after that. Table. \ref{tab:kripendorf} represents the intent values with their corresponding topic. Each sample in the collected interactive dataset has the form of $\xi_{j \dots n} \rightarrow \xi_{k \dots m}$, where there is no intersection between two sets of slots $\xi_{j \dots n} \cap \xi_{k \dots m} = \emptyset $. The selected slots are the slots the user selects during the interaction with the interactive agent. 
We collected more than $16,699$ user requests with $166,990$ user interactions for training, and $1,140$ user requests with $13,840$ interactions for testing.

\subsection{Corpus Collection}

To generate the corpus, we need to issue a series of queries to a search engine that will capture the search space of the web. Algorithm \ref{alg:subquery} shows the steps used to generate these queries and collect the corpus.

\begin{algorithm}
\caption{Algorithm to generate corpus for evaluation. \textbf{L} is the name of all US major cities. }
\label{alg:subquery}
\begin{algorithmic}
\State \textbf{Input: } $\Omega_{(\tau, i)}$, and \textbf{L} is the user location.
\State corpus = []
\For{each \textit{location} in database \textbf{L} }
\For{each $(\tau,i)$ in $\Omega$ } 
\For{each $\xi$ in $\Omega_{(\tau,i)}$ } 

\State query $\gets i + \text{near} + location + \text{with} + \xi $ 
\State top-100 documents $\gets$ search engine\_API (query) 
\State corpus = corpus $\cup$ top-100 documents
\EndFor
\EndFor
\EndFor
\State \textbf{Output: } corpus
\end{algorithmic}
\end{algorithm}

\section{Experimental Setup and Results}
\label{sec:experiements}

For convenience, we summarize the methods compared for reporting the experimental results as follows:

\subsubsection*{\bf Method 1: Popularity Method (Baseline)}
\label{sec:popularity}

The popularity-based method is a heuristic, suggesting the next set of related slots based on overall frequency (popularity) in the intent profile $\Phi$. The order of suggestions can change over time as some slots become more popular for specific intents.

\subsubsection*{\bf Group 1: Contextual Multi-armed Bandit Policies}
\label{sec:Bandits}
We report the results on $13$ different policies for contextual bandit models, including: ``Bootstrapped Upper Confidence Bound'', ``Bootstrapped Thompson Sampling'', ``Epsilon Greedy'', ``Adaptive Greedy'', ``SoftMax Explorer'', etc. which have been extensively investigated in~\citep{cortes2018adapting}. The library to implement the policies is available \href{https://contextual-bandits.readthedocs.io/en/latest/}{here}\footnote{\url{https://contextual-bandits.readthedocs.io/en/latest/}}. 
\subsubsection*{\bf Group 2: Different context representation:} We report the results for the three different proposed context $c$ described in Section~\ref{sec:interactive}.

\subsection{Evaluation Metrics}
\label{sec:eval_metrics}

Evaluating complex search tasks has always been quite challenging. Since the task is not supervised and there is no available dataset or labels, we could not directly evaluate the results. In addition, our goal is to refine requests in a way that they lead to better suggestions.  Therefore, we propose to employ \ac{QPP} metrics for evaluation purposes. 
\ac{QPP} task is defined as predicting the performance of a retrieval method on a given input request  \citep{carmel2010estimating,cronen2002predicting,he2004inferring}. In other words, query performance predictors predict the quality of retrieved items w.r.t to the query. QPP methods have been used in different applications such as query reformulation, query routing, and in intelligent systems~\citep{sarnikar2014query,roitman2019study}.
QPP methods are a promising indicator of retrieval performance and are categorized into pre-retrieval, and post-retrieval methods \cite{carmel2010estimating}.

Post-retrieval QPP methods generally show superior performance compared to pre-retrieval ones, whereas the pre-retrieval QPP methods have been more often used in more real-life applications and can address more practical problems since their prediction occurs before the retrieval. 
 
In addition, almost all of the post-retrieval methods work based on the relevance scores of the retrieved list of documents, and in our case, the relevance score was not available from the search engine API; thus, we only employed pre-retrieval QPP methods for this work’s evaluation purposes.
Having said that, we predict and compare the performance of the initial complex requests as well as our reformulated requests using SOTA pre-retrieval QPP methods which have been shown to have a high correlation with retrieval methods on different corpora \cite{hashemi2019performance,arabzadeh2020neural,arabzadeh2020neural1,zhao2008effective,hauff2008survey,hauff2009combination,carmel2012query,he2004inferring}.  The intuition behind evaluating our proposed method with pre-retrieval QPP methods is that QPP methods have shown to be a promising indicator of performance. Therefore, we can compare the predicted performance of the initial complex request as well as our reformulated request and predict which one is more likely to perform better. To simply put, higher \ac{QPP} values mean that there is a higher probability that the request is going to be easily satisfied, and lower \ac{QPP} values indicate a higher chance of poor retrieval results.

In the following, we elaborate on the SOTA pre-retrieval QPP methods that showed promising performance over different corpora and query sets, and we leveraged them for evaluating this work.

\textbf{Simplified Clarity Score (SCS):} SCS is a specificity-based QPP method, which captures the intuition behind that the more specific a query is, the more likely a system is to specify the query \cite{he2004inferring,plachouras2004university}. SCS measures the KL divergence between the query and the corpus language model, thereby capturing how well the query is distinguishable from the corpus.

\textbf{Similarity of Corpus and Query (SCQ): }
SCQ leverages the intuition that if a query is more similar to the collection, there is a higher potential to find an answer in the collection \cite{zhao2008effective}.
Concretely, the metric measures the similarity between collection and query for each term and then aggregates over the query, reporting the average of each query term's individual score. 

\textbf{Neural embedding based QPPs (Neural-CC):}
Neural embedding-based \ac{QPP} metrics have shown excellent performance on several \ac{IR} benchmarks. They go beyond the traditional term-frequency based \ac{QPP} metrics and capture the semantic aspects of terms \cite{zamani2018neural,roy2019estimating,arabzadeh2019geometric,arabzadeh2020neural,arabzadeh2020neural1,khodabakhsh2021semantics,roitman2020ictir,hashemi2019performance}. 
We adapted one of the recently proposed \ac{QPP} metrics which build a network between query terms and their most similar neighbors in the embedding space.
Similar to \citep{he2004inferring,plachouras2004university}, this metric is based on query specificity. The intuition behind this metric is that specific queries play a more central and vital role in their neighborhood network than more generic ones. Here, as suggested in~\citep{arabzadeh2020neural,arabzadeh2020neural1,arabzadeh2019geometric}, we adapted the Closeness Centrality (CC) of query terms within their neighborhood network, which has shown to have the highest correlation across different \ac{IR} benchmarks.

\begin{figure}
\centering
\includegraphics[width = 230pt]{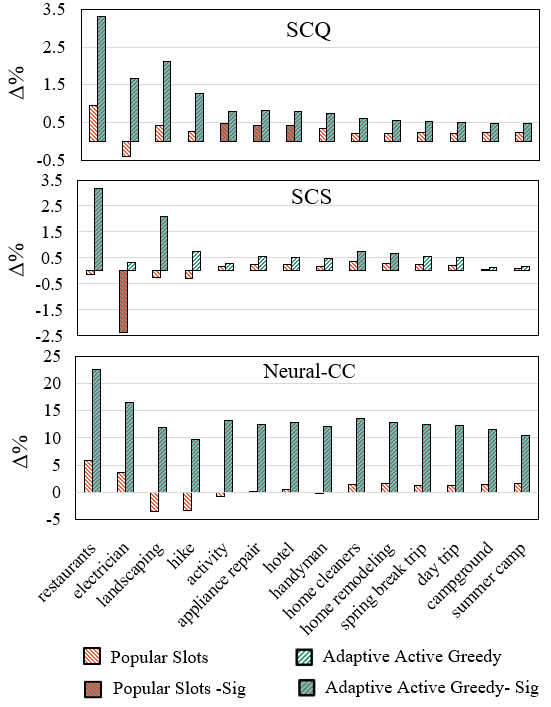}
\caption{\small Results of three \ac{QPP} metrics on reformulated queries difference percentile with original queries on all intents. The darker bars indicate statistically significant improvement with $\alpha=0.05$.}
\label{fig:mainresult}
\end{figure}

\subsubsection*{\bf Training Parameters:}
For contextual bandits and GPT-3 models, the default parameters for the available libraries were used, and no parameter tuning was performed. To train the deep learning model described in Section~\ref{sec:popularity}, we use an Adam optimizer with a learning rate of $\eta=0.001$, a mini-batch of size 8 for training, and embedding of size 100 for both words and aspects. 
A dropout rate of 0.5 is applied at the fully connected and ReLU layers to prevent potential overfitting. We used the default parameter for training; however, a smaller batch size was preferable based on the available dataset size. 

\subsection{Experimental Results}
\label{sec:exp_results}
We compare the result of \ac{QPP} metrics on our best policies and popular attributes with the original requests in Figure~\ref{fig:mainresult}, where we report the percentage difference w.r.t the full form. That is, to what extent do the QPP metrics predict that the reformulated requests are likely to perform better than the original ones. We examine the difference between the average of QPP metrics on reformulated requests with the best policy (adaptive active greedy) and the full form of requests. In addition, we compared the reformulated requests with popular attributes and the full form of the request and reported them in the same figure. As shown in Figure~\ref{fig:mainresult}, the adaptive active greedy policy showed improvements over all three QPP metrics and on all intents. 

These bars in Figure~\ref{fig:mainresult} can be interpreted as the percentage of predicted improvement for the reformulated requests compared to full form of requests. For instance, for restaurant intent, SCQ, SCS, and neural embeddings QPP methods, have improved by 3.3\%, 3.1\%, and 22.5\%, respectively.
We measure statistical significance using a paired t-test with a p-value of 0.05. We note that while the improvement made by the adaptive active greedy policy were consistently statistically significant on all intents by the SCQ and neural embedding QPP metrics, the gains were only statistically significant 4 intents on the SCS metric: ``Restaurants'', ``Landscaping'', ``Home cleaners'', and ``Home Remodeling.''  

 It should be noted that while QPP methods are potential indicators of performance, every QPP method focuses on a different quality aspect of the query. Therefore, they do not necessarily agree on the predicted performance according to different queries, corpora, or retrieval methods. This observation has been made on different SOTA QPP methods and various well-known corpora such as TREC corpora or MS MARCO and their associated query set~\cite{carmel2010estimating,arabzadeh2021bert,hashemi2019performance}. Thus, we conclude that the level of agreement could strengthen our confidence in the query performance prediction. In other words, the more QPP metrics agree on query performance, the more confidence we have in that prediction. In addition, we can interpret each QPP performance based on the intuition behind them. For example, the SCS method relies on query clarity level, while the SCQ method counts on the existence of potential answers in the corpus. When the two QPP methods do not agree on the query's performance, we consider it as how the query satisfies the intuition behind one of the QPP methods while failing to satisfy the others. For example, take the `activity' intent in Figure~\ref{fig:mainresult}, in which the SCQ methods showed significant improvement, but the SCS method did not. We interpret this observation as the clarity of the query has not been significantly increased while refined by our method. However, the query was expanded so that the existing potential answers in the corpus has increased.
 
\subsubsection*{\bf NLU evaluation}
To evaluate the topic and intent classifiers, the evaluation set described in Section~\ref{sec:datasetcollection} is used, which contains more than $16,699$ user requests with $166,990$ user interactions for training, and $1,140$ user requests with $13,840$ interactions for testing. The model achieved a 99.3\% and 95.2\% accuracy for topic and intents, respectively. For evaluating the slot tagger, we leveraged the annotated data collected by three different judges described in Section~\ref{sec:nludatacollection} performing 4-fold cross-validation and achieved a 0.75 macro-F1 across all the intents and slots. The results for slot tagging are promising despite the challenges, e.g., a small amount of labeled data, a large number of slots per intent, and overlapping slots across user intents. The results indicate the ability of GPT-3 to generalize in few-shot learning.

\subsection{Ablation Analysis}

\begin{figure}
\centering
\includegraphics[width = 210pt]{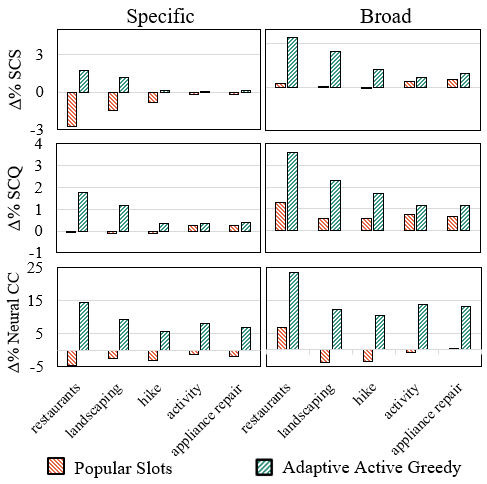}
\caption{\small Comparing Specific Requests vs Broad Requests in terms of 3 different pre-retrieval QPP metrics.}
\label{fig:broad-spec}
\end{figure} 

\subsubsection*{\bf Broad vs. Specific:} Studying a system's performance deeply on a per-query basis can enlighten where the systems fail, i.e., which queries a system fails to answer and which group of queries can be handled easily. Thus, it could potentially lead to future improvements to the system.
As such, exploring query characteristics has always attracted lots of attention between IR and NLP communities because query broadness has shown to be a crucial factor that could potentially lead to an unsatisfactory retrieval~\cite{song2009identification,clarke2009effectiveness,sanderson2008ambiguous,min2020ambigqa}. Here, we separately study the performance of our proposed method on two groups of broad and specific queries. We are interested in examining whether our proposed method can address both requests consistently, i.e., broad and specific ones. Here we define the \textit{broad requests} as the requests with less complex information-seeking tasks and fewer preferences expressed; they are short and contain a small number of slots/values ($\leq$ 3), hence requiring more steps for the \ac{RL} model to refine the user intent. On the contrary, the \textit{specific requests} is defined as the longer ones which contain many slots/values, and users need fewer steps to finalize their intent.

Figure~\ref{fig:broad-spec} presents the evaluation results of broad and specific requests. As demonstrated in this figure, although all the employed QPP metrics agreed that both types of requests had been improved, Adaptive Active Greedy would perform relatively better on broad queries compared to specific ones. It is an expected output because specific requests are more complex than broad ones, and more criteria should be addressed to satisfy them. Moreover, suggesting the popular slots have a deteriorating effect on all the metrics across the intents for the specific requests, showing a challenging reformulation process, while the proposed model in all metrics improves the QPP.

\begin{figure}
\centering
\includegraphics[width = 200pt]{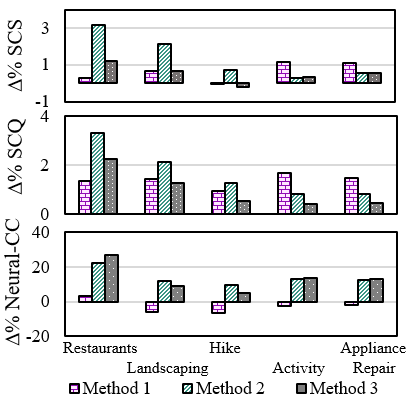}
\caption{\small Comparison analysis between different contexts w.r.t original form of requests based on pre-retrieval QPP metrics. }
\label{fig:dif-context}
\end{figure} 

\subsubsection*{\bf Different Context $c$:} We compare three different proposed contexts described in section \ref{sec:interactive} in terms of percentage difference with respect to the original form of requests on predicted performance by QPP metrics in Figure \ref{fig:dif-context} on the top-5 most popular intents. The results show all three proposed contexts outperform the original representation across all metrics and intents. We observe that QPP metrics do not consistently agree on the predicted performance between these three different methods. While neural-cc predicts that method 3 and method 2 to define the $c$ perform better than method 1. We also noticed that SCS and SCQ sometimes behave the opposite. We hypothesize that this difference could potentially be because neural-cc works based on neural embedding while SCS and SCQ work based on term frequency and corpus statistics. Therefore, each group might capture different aspects of requests. Although all the proposed contexts $c$ significantly outperform the original query, we cannot conclude which one among them outperforms the others.

\subsubsection*{\bf Policy evaluation for contextual bandit model: } We performed an experiment for policy evaluation on contextual bandits. We selected the popular intent for off-policy evaluation. Off-line contextual bandits assessment is complicated because they interact in online environments. There are multiple methods for policy evaluation for off-line settings such as Rejection Sampling (RS)~\citep{li2010contextual} and Normalised Capped Importance Sampling (NCIS)~\citep{gilotte2018offline}. All the results are reported based on the best arm performance. The system can expose users to multiple slots. As a result, in the proposed setting, the final performance will be much better than the described results. 

\begin{table}[t]
\small
\begin{center}
\begin{tabular}{l c c c cc}
\toprule
Avg. reward &restaurant&landscaping &hike&activity&appliance\\ 
\midrule
        RS   & 0.538  & 0.352   & 0.455& 0.25  &0.375\\   
        NCIS & 0.413  & 0.469   & 0.555& 0.407 &0.654\\
        Real & 0.378  & 0.440   & 0.495& 0.396& 0.670\\
\bottomrule

\end{tabular}
\end{center}
\caption{Policy evaluation results for RS and NCIS models.}
\label{tab:pram-tuning}
\end{table}

According to the results, RS sometimes underestimates the performance on intents like "restaurant" and ``appliance repair'' with overestimating other intents such as "hike." The NCIS method provides a more accurate estimation and provides a more realistic estimate.

\section{Discussion and Implications}
\label{sec:conclusion}
This paper proposed a novel application of natural language interfaces, allowing cold-start users to submit and receive responses for complex information-seeking requests expressed in natural language.

Unlike traditional search engines, where a single most relevant result is expected, users of our system are presented with a set of suggestions for further exploration.
We have designed and deployed a system that permitted us to conduct initial data collection and potential future online experimentation using the A/B testing paradigm.

To complement this platform for complex user requests, we leveraged the advances in generative language models and designed a platform to make them interactive for the future design of Search Engines. We developed a novel interactive agent based on contextual bandits that guide users to express the initial request more clearly by refining their intents and adding new preferential desiderata. During this guided interaction, a \ac{NLU} unit, designed on top of an LLM model, is used to build a structured semantic representation of the request. The system also uses a proposed \acf{CIS} that estimates the degree to which intent is entirely expressed at each interaction step.

To efficiently leverage the power of generative language models in designing NLU, we used the few-shot prompting technique. We proposed a pipeline named Pluto to collect high-quality samples labeled by human annotators described in section \label{sec:nludatacollection}. Pluto used a human-in-the-loop setup for data collection and consisted of two main components 1) Refinement of complex user requests in natural language, and 2)
Refinement of the retrieved list of suggestions. The training set along with the current user request is used to form the prompt for GPT-3, then the extracted aspects were used as an initial point for an exploratory search to elicit user preferences.

These high-quality samples are used as a few-shot prompt and concatenated with the final user query to form the final prompt for the NLU inference.

When the system determines that an optimal request has been expressed, it leverages a search API to retrieve a list of suggestions. To demonstrate the efficacy of the proposed modeling paradigm we have adopted various pre-retrieval metrics that capture the extent to which guided interactions with our system yield better retrieval results. In a suite of experiments, we demonstrated that our method significantly outperforms several robust baseline approaches. We used three SOTA pre-retrieval QPPP metrics such as SCQ, SCS, and neural embeddings QPP to evaluate our method.  The proposed interactive LLM model had promising results on different user intents e.g., restaurant intent, where SCQ, SCS, and neural embeddings QPP improved by 3.3\%, 3.1\%, and 22.5\%, respectively.

This article focused on making generative language models interactive and alleviating their problem in responding to immediate user complex information-seeking problems. The generative language models suffer from hallucinating in providing accurate factual knowledge, especially user questions about what is going on in the world currently, as nearly all of them are trained on huge amounts of data collected in the past. The proposed IA can also benefit from the fast advancements in the generative language models, such as ChatGPT \cite{leiter2023chatgpt} and GPT-4 \cite{gpt4}, and replace its core NLU component with the larger and more advanced models.

In future work, we plan to design an online experiment that will involve business metrics, such as user satisfaction and the ratio of returning users and interactively collect ratings for the list of suggestions made by our system. This will allow us to learn from language and rating data jointly. Another possible direction is designing intent ontologies in a more complex hierarchical form where there are more complex and hierarchical dependencies between attributes. Finally we plan to investigate the reliance of the proposed interactive LLM model on GPT-3 as compared to the more recent larger models such as GPT-4, specifically the study of current drawbacks of current LLM such as  hallucination and its impact on the proposed model. 

\bibliography{bibliography}
\bibliographystyle{plainnat}

\end{document}